# Topographic and electronic structure of cleaved SrTiO$_3$(001) surfaces

Wattaka Sitaputra[a), b)]

Carnegie Mellon University, Department of Materials Science and Engineering, Robert Engineering Hall, 5000 Forbes Avenue, Pittsburgh, Pennsylvania 15213

Marek Skowronski

Carnegie Mellon University, Department of Materials Science and Engineering, Robert Engineering Hall, 5000 Forbes Avenue, Pittsburgh, Pennsylvania 15213

Randall M. Feenstra[a)]

Carnegie Mellon University, Department of Physics, Wean Hall, 5000 Forbes Avenue, Pittsburgh, Pennsylvania 15213

[a)]American Vacuum Society member.

[b)]Electronic mail: wattaka@andrew.cmu.edu

The topographic and electronic structure of cleaved SrTiO$_3$(001) surfaces were studied, employing samples that either had or had not been coated with Ti on their outer surfaces prior to fracture. In both cases, SrO- and TiO$_2$-terminated terraces were present on the cleavage surface, enabling *in situ* studies on either termination. However, the samples coated with Ti prior to fracture were found to yield a rougher morphology on TiO$_2$-terminated terraces as well as a higher density of oxygen vacancies during an annealing (outgassing) step following the coating. The higher density of oxygen vacancies in the bulk of the Ti-coated samples also provides higher conductivity which, in turn, improves a sensitivity of the spectroscopy and reduces the effect of tip-induced band bending. Nonetheless, similar spectral features, unique to each termination, were observed for samples both with and without the Ti coating. Notably, with moderate-temperature annealing following fracture, a strong discrete peak in the conductance spectra, arising from oxygen vacancies, was observed on the SrO-terminated terraces. This peak appears at slightly different voltages for coated and uncoated samples, signifying a possible effect of tip-induced band bending.

## I. INTRODUCTION

Cleaved surfaces have been routinely used as a platform for scanning tunneling microscopy (STM) and spectroscopy (STS) studies as they provide a high-quality adsorbate-free surface with large terraces given that the material of interest has a natural cleavage plane.[1,2] Examples of materials with the natural cleavage plane can be found among elemental and compound semiconductors (Si, Ge, GaAs, InSb, etc).[3–7] $SrTiO_3$, on the other hand, does not have a natural cleavage plane and fracture typically produces a rough surface unsuitable for STM and STS. However, by implementing extra care during the cleaving process, a macroscopically rough yet nanoscopically smooth surfaces can be obtained, as previously demonstrated by Guisinger *et al*.[8]

The fractured surfaces of $SrTiO_3$ have become increasingly interesting owing to the discovery made by Santander-Syro *et al*.[9] It was found that, by cleaving the $SrTiO_3$ in ultra-high vacuum at low temperature along a (100) plane and exposing the surface to intense ultraviolet (UV) light, one can generate a 2-dimensional electron gas (2DEG) on the surface. Oxygen vacancies are believed to be created during the fracture/exposure and to serve as source of electrons for the 2DEG.[9,10] However, to date, the formation of the 2DEG is limited to low temperature cleaved surfaces, since room temperature cleaves produce a higher defect density and a smaller terrace widths.[11,12]

In order to better understand a defective nature of surfaces cleaved at room temperature, we have previously explored the electronic structure of such surfaces using STM/STS, observing features arising from oxygen vacancies on the surface as well as disorder-induced surface states, and the effects of moderate-temperature annealing.[13] In these experiments the annealing procedure (both after fracture, and before fracture when the annealing is employed as an outgassing step) was performed by passing current through the sample, i.e. from front to back of a thin sample. This method generally produces a high temperature gradient (or a hot spot), with the location depending on mechanical contacts between the sample and electrical leads. These hot spots introduce nonuniformity in the local density of oxygen vacancies formed by the annealing. In order to circumvent this problem, a layer of titanium (Ti) was deposited on both polished surfaces of the sample prior to fracture, and improved uniformity during heating was thus achieved. However, variations in both topographic structure and spectroscopy of the cleaved surfaces were observed as a result of the Ti coating, and these variations require further investigation.

In this work, we have studied the development in topographic and electronic structure before and after moderate-temperature annealing of the Ti-coated and uncoated samples, in order to better understand the effects of Ti coating and ensure the consistency of the spectroscopic results. It was found that the Ti coating does not affect the unique characteristics of the cleaved $SrTiO_3$(001) surfaces, i.e. stepped terraces with alternating SrO and $TiO_2$ termination and their associated spectral features. However, the Ti coating produces a remarkably different surface morphology for $TiO_2$-terminated terraces, and furthermore it increases a density of oxygen vacancies due to annealing (as one would expect from a reaction between Ti and $SrTiO_3$).[14] This latter effect turns out to be favorable for the experiments, since the increased density improves the bulk conductivity and hence facilitates the spectroscopic observations and analysis.

## II. EXPERIMENTAL

Cleaved surfaces of $SrTiO_3$(001) were prepared by fracturing 0.05 wt% Nb-doped $SrTiO_3$(100) single crystals along a (100) plane in ultra-high vacuum (UHV), at room temperature. Prior to cleaving, most of samples were sputter coated with 100 nm of titanium on both polished surfaces. During this step, a residual metallic layer on the side of the sample was carefully removed in order to ensure that a current will pass through sample, not along the surface, when a bias is applied across each surface during resistive heating, as shown in Fig. 1(b). Advantages of this procedure include optimum heating efficiency, uniform power dissipation and effective reduction of the $SrTiO_3$ samples, which in turn produce a uniform distribution of oxygen vacancies during the outgassing step (described below). After the deposition of the titanium, a groove along a <100> direction was cut using a diamond wire saw, with a depth approximately 60-70% of the total thickness of the sample (Fig. 1).

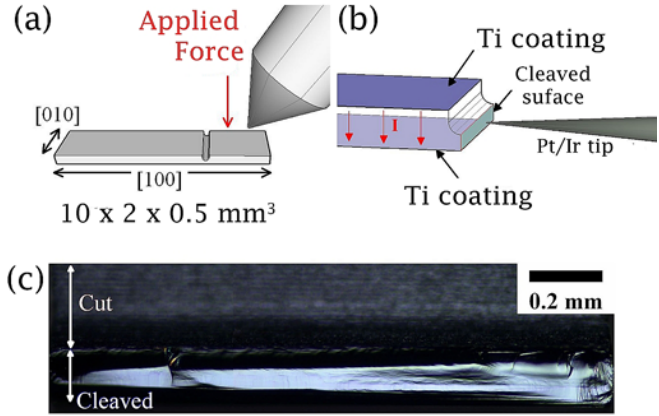

**FIG. 1** (Color online) (a) Schematic diagram illustrating sample geometry. (b) Configuration for tunneling and heating, with "I" representing a direction of the current. The Ti coating on the upper and lower surfaces of the sample is indicated. (c). Optical micrograph of a cleaved $SrTiO_3$(100) surface. The relatively thick "cut" region is indicated, as is the narrower "cleaved" region. The latter is bright (due to an optical reflection) where the fractured surface is nearly parallel to the (100) plane.

After preparation, the samples were mounted onto a sample holder with the groove placed at the edge of the sample holder. The samples were then loaded into the UHV chamber with a base pressure of $4 \times 10^{-11}$ Torr and outgassed by resistive heating at 700-800°C for 5 minutes. The outgassing step not only provides a cleaner environment for tunneling but also creates a high density of oxygen vacancies throughout the volume of the crystal. The final step was to cleave the sample at room temperature by pushing the portion of sample protruding beyond the sample holder.

STM and STS were performed at room temperature with a Pt/Ir tip. A tunnel current of 0.1 nA and a sample bias in a range of 1.25 - 3.00 V were used for acquisition of topographic images and conductance maps. A lock-in technique was used to obtain differential tunneling conductance (dI/dV) spectra with oscillation frequency of 1 kHz and rms modulation amplitude of 50 mV. A ramp in the tip-sample separation as a function of bias voltage (1.5 Å/V) was employed, as described by by Mårtensson and Feenstra.[15] This technique provides a higher sensitivity to the spectroscopy and enables us to obtain data with a dynamic range of 4 - 5 orders of magnitude.[13]

## III. RESULTS
### A. Cleaved surfaces without Ti coating

As illustrated in Fig. 2(a) and (b), cleaved (100) surfaces of SrTiO$_3$ reveal a stepped-terrace structure and a conductance stripe pattern, signifying alternating terminations.[8] The roughness of the two types of terminations is comparable for our samples that are not coated with Ti, and this morphology is similar to that observed by Guisinger *et al*.[8] A step height which is multiple of a unit cell (3.905 Å) high was observed between planes with the same termination (Fig 2(c)). However, the step height between planes with different termination was found to be slightly smaller than expected, i.e. less than half a unit cell high (or 1.953 Å). This discrepancy can be understood by taking into account the difference in total conductance (I/V) between the two terminations, at the particular sample bias employed. For instance, the TiO$_2$-terminated terraces exhibit lower conductance than the SrO-terminated terraces at a sample bias of +3.0 V, so that in order to maintain the same tunnel current the tip has to move closer to the surface when it crosses over from the SrO-terminated terrace onto the TiO$_2$-terminated terrace, as seen in Fig. 2(c). The map of differential conductance (dI/dV), Fig. 2(c), reveals contrast in accordance to the spectra shown in Fig. 2(d), with the SrO-terminated terraces appearing brighter. It is important to note that this change in apparent step height as well as the contrast in conductance map will vary with bias voltage due to the difference in electronic structure between the SrO and TiO$_2$ terminations (Fig. 2(d)).

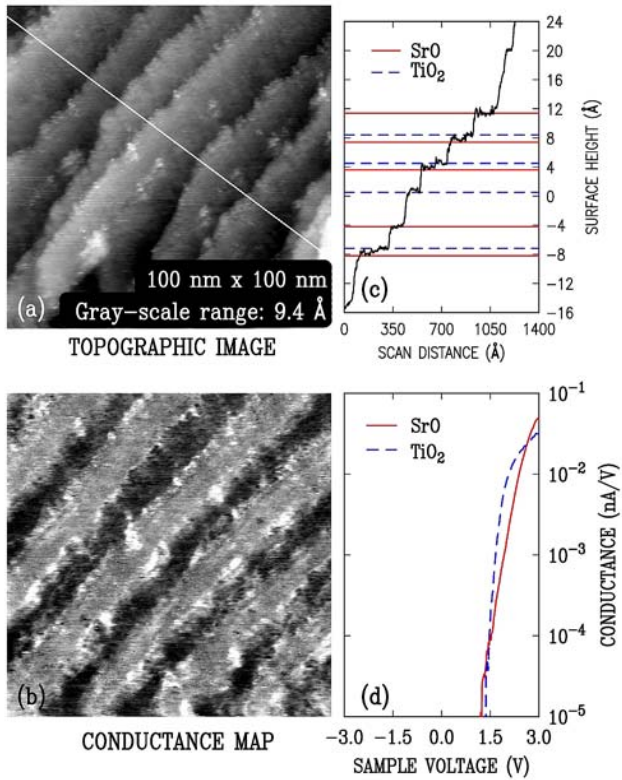

**FIG. 2** (Color online) (a) Topographic image and (b) conductance map of the cleaved SrTiO$_3$(100) surface obtained with a sample bias of +3.0 V. At this bias, bright conductance stripes occur for SrO termination while dark stripes occur for TiO$_2$ termination. (c) A height profile obtained along the specified path in (a). Red solid lines show SrO-terminated terraces while blue dashed lines show TiO$_2$-terminated terraces. (d) Conductance versus voltage spectra averaged across their respective SrO- and TiO$_2$-

terminated terraces and plotted in logarithmic scale. The Fermi level occurs at 0 V. Note: no titanium coating was performed on this sample.

As a result, to reliably identify a change in surface termination, one should look for a unique feature for each termination within a conductance spectra which is highly reproducible even under the presence of varying surface states. It was reported by Guisinger *et al.*[8] that a peak or a shoulder near +2.25 V signifies the $TiO_2$ termination while a monotonic increase near +3.0 V signifies the SrO termination. We also found these features, shown in Fig. 2(d), to be highly reproducible for both cleaved surfaces and epitaxially grown surfaces, the latter reported elsewhere.[13]

It is also worth mentioning that, for our uncoated samples, a valence band (VB) edge is completely absent in the spectra for both terminations, for sample voltages as low as -3.0 V. This result is consistent with the prior report of Guisinger *et al.*[8] The absence of the VB edge can be generally attributed to a combined effect of equilibrium tip-induced band bending as well as nonequilibrium band bending resulting from surface charging.[16] It is likely that the latter effect dominates in this case since the bulk conductivity of our samples is not so low (~0.1 $ohm^{-1}cm^{-1}$ for the as-received wafers, and probably very much higher following the annealing steps). The situation for surface charging at negative sample voltages would be similar to that described in Fig. 7 of Ref. [16]: positive surface charge is produced by depletion of the surface donor states, causing a reduction in band bending until flat band conditions are reached. Flat band conditions would be achieved for a sample voltage approximately equal to the band gap. Although the spectra displayed here are limited on the negative side to -3.0 V, we have occasionally measured the current for larger negative voltages and we do indeed observe a sharp increase in the current at voltages slightly beyond -3.0 V, consistent with this surface charging argument. In any case, as discussed below, with increased numbers of oxygen vacancies (and hence increased bulk and surface conductivity, which reduce the non-equilibrium surface charging effects), the VB edge become visible in the spectra.

## B. *Cleaved surfaces with Ti coating*

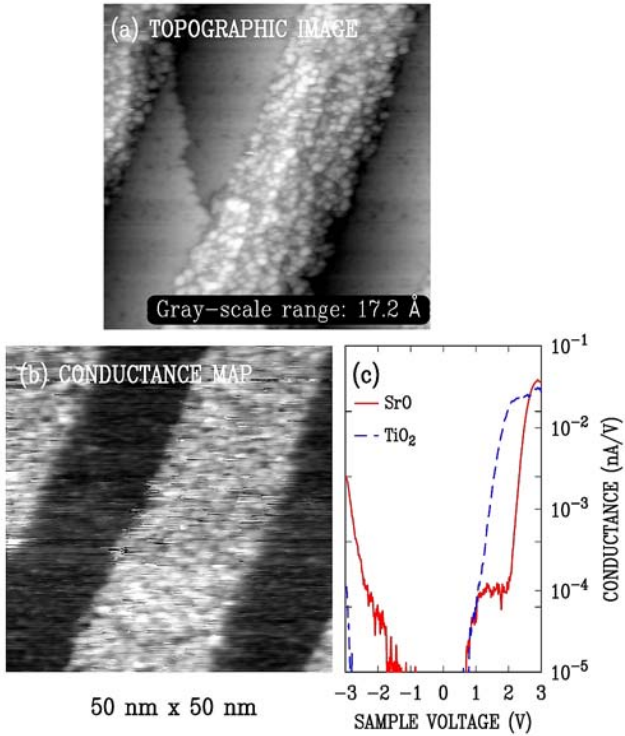

**FIG. 3** (Color online) (a) Topographic image, and (b) conductance map of the cleaved $SrTiO_3(100)$ surface obtained with a sample bias of +1.5 V. At this bias, bright conductance stripes occur for $TiO_2$ termination while dark stripes occur for SrO termination. (c) Conductance versus voltage spectra averaged across their respective SrO- and $TiO_2$-terminated terraces. The Fermi level occurs at 0 V. Note: titanium coating was used on this sample.

In contrast to the situation for samples with no Ti coating prior to cleavage, in which the topographic structure of the SrO- and $TiO_2$-terminated surfaces are similar, we find for samples which are coated with Ti that a distinct difference in topographic structure occurs between the two terminations. The $TiO_2$-terminated surfaces are now significantly rougher than the SrO-terminated ones (Fig. 3(a) and (b)). The roughness of the $TiO_2$-terminated terraces is also significantly greater than what was observed on the uncoated samples (Fig. 2(a)). It is important to note that this result differs with previous work in which flatter terraces were generally associated with $TiO_2$ termination.[8,11] Apparently, the differing strain state of the sample during cleavage due to the Ti coating (and/or a greater number of bulk vacancies) affects the fracture mechanism. Nevertheless, all important spectral characteristics such as a shoulder near +2 V for $TiO_2$ termination, a monotonic increase near +3 V for SrO termination and stripe pattern in conductance map are clearly preserved as shown in Fig. 3(b) and (c). Thus, different surface terminations can still be consistently differentiated and identified.

Along with a change in topographic structure, additional features in conductance spectra also emerge on the Ti-coated samples (Fig. 3(c)). As the samples with Ti coating produce a considerably greater number of oxygen vacancies, a transition level from neutral to +1 charged state of oxygen vacancy can now be observed on SrO-terminated terraces as a small shoulder near +1.3 V.[13] In addition, these oxygen vacancies, which act as donors, also improve the conductivity throughout the sample and, thus, improve the capability for the sample to conduct the tunnel charges away. As a result, the sensitivity of the spectroscopy for VB states is improved and the band bending effect is reduced. Consequently, a VB edge can now be observed (Fig. 3(c)).

## C.  Effects of moderate-temperature annealing

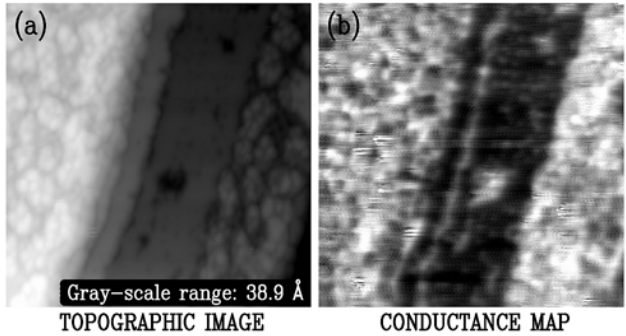

**FIG. 4** (a) Topographic image and (b) conductance map, acquired with a sample bias of +1.75 V and extend over 50 nm x 50 nm, after 10 minutes of annealing at 360ºC. Note: titanium coating was used on this sample.

Now let us turn to detailed discussion of annealing at moderate temperatures. *In situ* annealing in the temperature range of 260-360ºC for 10 minutes is found to produce restructuring of the $TiO_2$ surface plane for the Ti-coated samples: the surface which is initially rough on an atomic scale develops topography with distinct, atomically flat terraces separated by steps (Fig. 3(a) and Fig 4(a)). The SrO-terminated terraces remained unchanged. The restructuring of the $TiO_2$-terminated surfaces confirms that surface atoms are mobile at these temperatures. Additionally, changes in electronic structure are found on both the $TiO_2$-terminated and the SrO-terminated surfaces, indicating that oxygen vacancies are also mobile in this temperature range. This movement of oxygen vacancies is reasonable since they are expected to be mobile even at room temperature.[17]

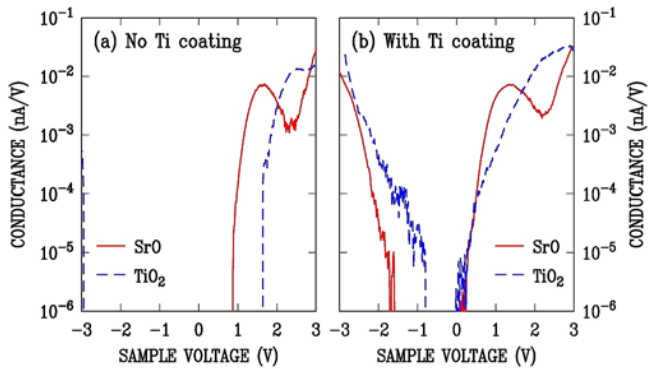

**FIG. 5** (Color online) Averaged conductance spectra acquired from each termination after moderate temperature annealing of (a) uncoated samples and (b) coated samples.

By segregating oxygen vacancies onto the surface with the moderate-temperature annealing, a discrete peak located at +1.6 V for uncoated samples and +1.3 V for Ti-coated samples is found to be produced on SrO-terminated terraces (Fig. 5). We have argued elsewhere that this peak arises from surface oxygen vacancies[13] and it can be regarded as an enhanced form of the shoulder at +1.3 V in Fig. 3(c). Notably, the oxygen vacancy peak tends to be reproducible at any location on the SrO-terminated surfaces. The variation in probing position is found to only affect the magnitude of this in-gap feature, not its position in the spectrum, i.e. there is no consistent shift in peak position as the tip moves toward the center of the terrace.

The vacancy peak position shifts by about 0.3 V between coated and uncoated samples (Fig. 5). The shift signifies slightly different band bending between the two samples; careful inspection of the CB onset for the $TiO_2$ termination reveals a similar shift. This upward shifting of spectral features that occurs for the sample without Ti coating arises, we believe, most likely from band bending due to surface charging, similar to that discussed at the end of Section III(A) above. Again referring to Fig. 7 of Ref. [16], at positive voltages some nonequilibrium occupation of surface acceptors (produced by surface disorder) will occur, producing the upwards band bending. For the Ti-coated samples, a much higher concentration of oxygen vacancies is likely produced by reduction of the sample by the Ti film, hence resulting in higher surface and bulk conductivity and occupations of the surface acceptors that are closer to equilibrium. We note that the observed band gap in the spectra of the Ti-coated samples is close to the bulk gap for $SrTiO_3$,[13] consistent with this picture of only small amounts of equilibrium tip-induced band bending in our experiments.

With regard to disorder-induced surface states, the extent of its effect can be clearly seen in a conductance spectrum of $TiO_2$-terminated surfaces after annealing (Fig. 5(b)). A dramatic increase in conductance was observed on a negative-voltage side, together with a tail of states extending toward the Fermi level from the shoulder at +2 V. We believe this change is associated with a change in topographic structure as shown in Fig. 4(a) (in comparison to Fig. 3(a)). As it turns out, the formation of a more distinct stepped-terrace structure results in an increasing density of disorder-induced surface states which appear inside the band gap, although without any discrete features or peaks appearing in this spectrum of states.

## IV. SUMMARY AND CONCLUSIONS

In summary, we have discussed differences in the topographic and electronic structure of cleaved $SrTiO_3$ surfaces with and without Ti coating prior to fracturing. Without Ti coating, the cleaved surfaces exhibit comparable roughness for both $TiO_2$ and SrO termination. However, when the sample is Ti coated before cleavage, the $TiO_2$-termianted terraces were found to be substantially rougher. This change in roughness can be attributed to a difference in strain state as well as a difference in density of oxygen vacancies formed during the outgassing step. Even with this difference, the spectral characteristics unique to each termination are maintained for both coated and uncoated samples. However, there is a remarkable difference in the observed tunneling spectra on the negative voltage side between the two types of samples. The VB states were found to be almost completely absent for the uncoated samples, which is attributed to their lower density of oxygen vacancies, thus, lower bulk and surface conductivity. Even after moderate temperature annealing, which serves to segregate oxygen vacancies toward the surface, the VB states are still absent for the uncoated sample, confirming the importance of bulk/surface conductivity on the visibility of the VB states of our $SrTiO_3$ sample. It is also owing to this insufficient conductivity that the position of the oxygen vacancy transition level is slightly shifted in spectra obtained from the uncoated samples. The VB states for the samples coated with Ti, on the other hand, are clearly visible. A drastic change in disorder-induced surface states was also observed on the $TiO_2$-terminated surfaces, which is attributed to observed surface restructuring after annealing.

## ACKNOWLEDGMENTS

We thank Mohamed Abdelmoula and Ying Lu for their help with titanium deposition. This research was supported by AFOSR Grant No. FA9550-12-1-0479.